\documentclass[useAMS,usenatbib]{mnras}
\usepackage{graphicx}
\usepackage{ulem}
\usepackage{amsmath}
\usepackage{amssymb}

\def\ltsima{$\; \buildrel < \over \sim \;$}
\def\simlt{\lower.5ex\hbox{\ltsima}}   
\def\gtsima{$\; \buildrel > \over \sim \;$}
\def\simgt{\lower.5ex\hbox{\gtsima}}

\def\EMCEE{{\sc Emcee}}

\def\mathnew{\mathsurround=0pt}   
\def\simov#1#2{\lower .5pt\vbox{\baselineskip0pt  
    \lineskip-.5pt\ialign{$\mathnew#1\hfil##\hfil$\crcr#2\crcr\sim\crcr}}}

\def\'#1{\ifx#1i{\accent"13\i}\else{\accent"13#1}\fi}

\title[An efficient positive potential-density pair expansion for modelling galaxies
]{An efficient positive potential-density pair expansion for modelling galaxies}

\author[A. Rojas-Ni\~no, J. I. Read, L. Aguilar \& M. Delorme]{A. Rojas-Ni\~no$^{1}$\thanks{Contact e-mail: armando.rojas@iems.edu.mx}\thanks{Present address: Instituto de Educaci\'on Media Superior del Distrito Federal, Av. Divisi\'on del Norte 906, C.P. 03020, M\'exico D.F.}, J. I. Read$^{1}$, L. Aguilar$^{2}$ and M. Delorme$^{1}$\\
$^1${\small Department of Physics, University of Surrey, Guildford, GU2 7XH, Surrey, UK}\\
$^2${\small Instituto de Astronom\'ia, Universidad Nacional Aut\'onoma de M\'exico, Apartado Postal 877, 22860 Ensenada, B.C., M\'exico} 
}

\begin{document}

\maketitle

\begin{abstract}
We present a novel positive potential-density pair expansion for modelling galaxies, based on the Miyamoto-Nagai (MN) disc. By using three sets of such discs, each one of them aligned along each symmetry axis, we are able to reconstruct a broad range of potentials that correspond to density profiles from exponential discs to 3D power law models with varying triaxiality (henceforth simply ``twisted" models). We increase the efficiency of our expansion by allowing the scale length parameter of each disc to be negative. We show that, for suitable priors on the scale length and height parameters, these ``MNn discs" have just one negative density minimum. This allows us to ensure global positivity by demanding that the total density at the global minimum is positive. We find that at better than 10\% accuracy in our density reconstruction, we can represent a radial and vertical exponential disc over $0.1-10$ scale lengths/heights with 4 MNn discs; an NFW profile over $0.1-10$ scale lengths with 4 MNn discs; and a twisted triaxial NFW profile with 3 MNn discs per symmetry axis. Our expansion is efficient, fully analytic, and well-suited to reproducing the density distribution and gravitational potential of galaxies from discs to ellipsoids.
\end{abstract}


\section{Introduction}
\label{sec:intro}
Building mass models of galaxies has a long and rich history dating back to the pioneering work of \citet{1922ApJ....55..302K} and \citet{1956BAN..13..15}. Modern models typically decompose disc galaxies into a stellar bulge and disc and a spherical or near-spherical dark matter halo \citep[e.g.][]{1981ApJ...251...61C,1983ApJ...265..730B,1991RMxAA..22..255A,DehnenBinney1998,2000MNRAS.316..929E,1999MNRAS.310..645W,2000MNRAS.311..441C,2001MNRAS.323..285B,GentileEtal2004,2011MNRAS.414.2446M}; while elliptical galaxies are modelled as axisymmetric or triaxial systems \citep[e.g.][]{1978ComAp...8...27B,1982ApJ...263..599S,1987ApJ...314..476L,1999MNRAS.303..455M,2004MNRAS.353....1S,2007MNRAS.379..418C,2013MNRAS.435.3587F}. Such modelling provides constraints on galaxy formation theories \citep[e.g.][]{2011MNRAS.416..322D,2013MNRAS.432.1862C,2015MNRAS.453.2447D}; the nature of dark matter \citep[e.g.][]{1959BAN....14..323V,1970ApJ...159..379R,Freeman1970,1980ApJ...238..471R,1985ApJ...295..305V,2005MNRAS.361..971R}; the initial mass function of stars \citep[e.g.][]{2015MNRAS.446..493P}; and even the presence of central supermassive black holes \citep[e.g.][]{1998AJ....115.2285M}.

Typically, mass modelling practitioners assume some parameterized form for the gravitational potential of the stellar and dark components. This provides a convenient minimal description, but will lead to systematic bias if the chosen model does not encompass the object being fitted. This has led to the development of `free form' or non-parametric methods \citep[e.g.][]{2001AJ....122..232C,2013ApJ...765L..15I,2013MNRAS.436.2598W,2015AJ....149..180O,2015arXiv150708581S}. These assume some rather general expression for the potential, typically with far more parameters than data constraints. These parameters are then marginalised over using statistical techniques like Markov Chain Monte Carlo \citep[e.g.][]{2013ApJ...765L..15I,2013MNRAS.436.2598W,2014arXiv1401.1052C,2015arXiv150708581S}. At the extremum of such methods is an expansion of the potential with arbitrarily many terms. \citet{1992ApJ...386..375H} refer to this method of solving the Poisson equation as `Self Consistent Field', or SCF; it can be made efficient if the lowest order terms in the expansion reasonably approximate real galaxies \citep{1973Ap&SS..23...55C,1992ApJ...386..375H}. 

With the advent of exquisite data from integral field spectrographs and the Gaia satellite, we can now afford to relax many of the traditional assumptions that go into the mass models \citep[e.g.][]{2011MNRAS.413..813C,2013arXiv1310.3485B,2015arXiv150708581S}. To this end, it is interesting to explore new expansions for the gravitational potential. While an orthonormal and complete basis is ideal, there is no guarantee that this will lead to a positive mass distribution everywhere in space. Furthermore, such fully general expansions are often not very efficient, being typically either well suited to a disc-like geometry or a more spheroidal geometry \citep[e.g.][]{1992ApJ...386..375H}. 

An alternative to a complete basis function is to utilise some general form for the gravitational potential $\Phi(x,y,z)$ or density $\rho(x,y,z)$, chosen to approximate real galaxies. If using an analytic form for the potential, then the density follows straightforwardly from Poisson's equation:

\begin{equation}
\nabla^2 \Phi = 4\pi G \rho.
\label{eqn:poisson}
\end{equation}
However, the trouble with this approach is that it is difficult to ensure that $\rho$ will be everywhere positive. A simple example is the flattened logarithmic potential for which the density is negative for flattening parameter $q < 1/ \sqrt{2}$ \citep{1987gady.book.....B}. If instead, we use some general form for the density distribution $\rho(x,y,z)$, then solving the Poisson equation becomes a potentially expensive inversion problem.

In this paper, we take an approach similar to the `Multi-Gaussian Expansion' method \citep{2002MNRAS.333..400C} that has been widely used for modelling galaxies and globular star clusters \citep [e.g.][]{2013MNRAS.436.2598W,2013MNRAS.432.1862C}. However, instead of using Gaussians that require a one-dimensional integral to determine the gravitational potential, we use instead an expansion in known analytic `potential-density pairs'. These are known analytic solutions to the Poisson equation and so $\Phi(x,y,z)$ is fully specified by $\rho(x,y,z)$ and vice-versa. Specifically, we focus on the Miyamoto-Nagai (MN) density profile which has already been widely used to model galactic discs \citep[e.g.][]{1975PASJ...27..533M,1981ApJ...251...61C,1991RMxAA..22..255A,1996MNRAS.281.1027F,2005MNRAS.361..971R,2015MNRAS.448.2934S}. However, our method could equally be applied to any potential density-pair. The MN density profile and potential are given by:

\begin{eqnarray}
\rho_{\rm MN}(R,z) = \hspace{3.5cm} \nonumber \\ 
\hspace{0.3cm} {{b^2M}[{aR^2+(a+3\sqrt{z^2+b^2})(a+\sqrt{z^2+b^2})^2}] \over {4\pi([R^2+(a+\sqrt{z^2+b^2})^2]^{5/2}{(z^2+b^2)}^{3/2})}}.
\label{DenMiya}
\end{eqnarray}
\begin{equation}
\Phi_{\rm MN}(R,z) = -{GM \over \sqrt{R^2+(a+\sqrt{z^2+b^2})^2}},
\label{PotMiya}
\end{equation}
where $R,z$ are the radius and height in cylindrical coordinates; $M$ is the disc mass and $a$ and $b$ play the role of the radial and height scale length of the disc respectively. Notice that in the limit $a \rightarrow 0$, the MN disc approaches the Plummer model \citep{1911MNRAS..71..460P}; while in the limit $b \rightarrow 0$ it approaches the infinitely thin Kuzmin disc \citep{Kuzmin1956}. 

The MN disc is an attractive building block for a new expansion function since any linear combination of such discs will provide an analytic and positive solution to the Poisson equation for all values of $M, a, b > 0$. Indeed, the early `Satoh models' took advantage of this property \citet{1980PASJ...32...41S}. Furthermore, a single component can already approximate both disc-like structures and spherical/oblate Plummer-like distributions using just three parameters (two length parameters and a mass parameter). Recently, \citet{2015MNRAS.448.2934S} have shown that three MN discs give a good fit to an exponential disc (that is a good approximation of the stellar disc of both our Milky Way and many extragalactic systems; e.g. \citealt{1987ApJ...320L..87L,2006A&A...454..759P,2013A&ARv..21...61R}) out to 10 disc scale lengths. In their model, the MN discs are allowed to have negative mass which is why so few terms are required to give an excellent fit. However, the price for this is that the density becomes negative at large radii. In this paper, we expand on this idea by introducing three new ingredients. Firstly, we do not allow negative mass $M$ but we allow instead the parameter $a$ to go negative. We refer to these as `Miyamoto-Nagai negative' (MNn) discs. We show that with suitable priors on the parameters $a$ and $b$, global minima must lie along $R$ in the $z=0$ plane. Thus, by performing a rapid test of positivity along this line, we can efficiently ensure that our entire reconstruction is positive. Secondly, we align three sets of such discs along each symmetry plane allowing us to build fully analytic triaxial models. Thirdly, we allow arbitrarily many terms in our expansion, allowing us to reach high accuracy where the data require it.

This paper is organised as follows. In \S\ref{method}, we describe our MNn disc expansion method. In \S\ref{sec:results}, we present some worked examples ranging from exponential discs to twisted triaxial ellipsoids. Finally in \S\ref{conclusions}, we present our conclusions.

\section{The method: the superposition of Miyamoto-Nagai discs}\label{method} 

\subsection{The Miyamoto-Nagai `negative' (MNn) disc}

\begin{figure}
\includegraphics[width=0.5\textwidth]{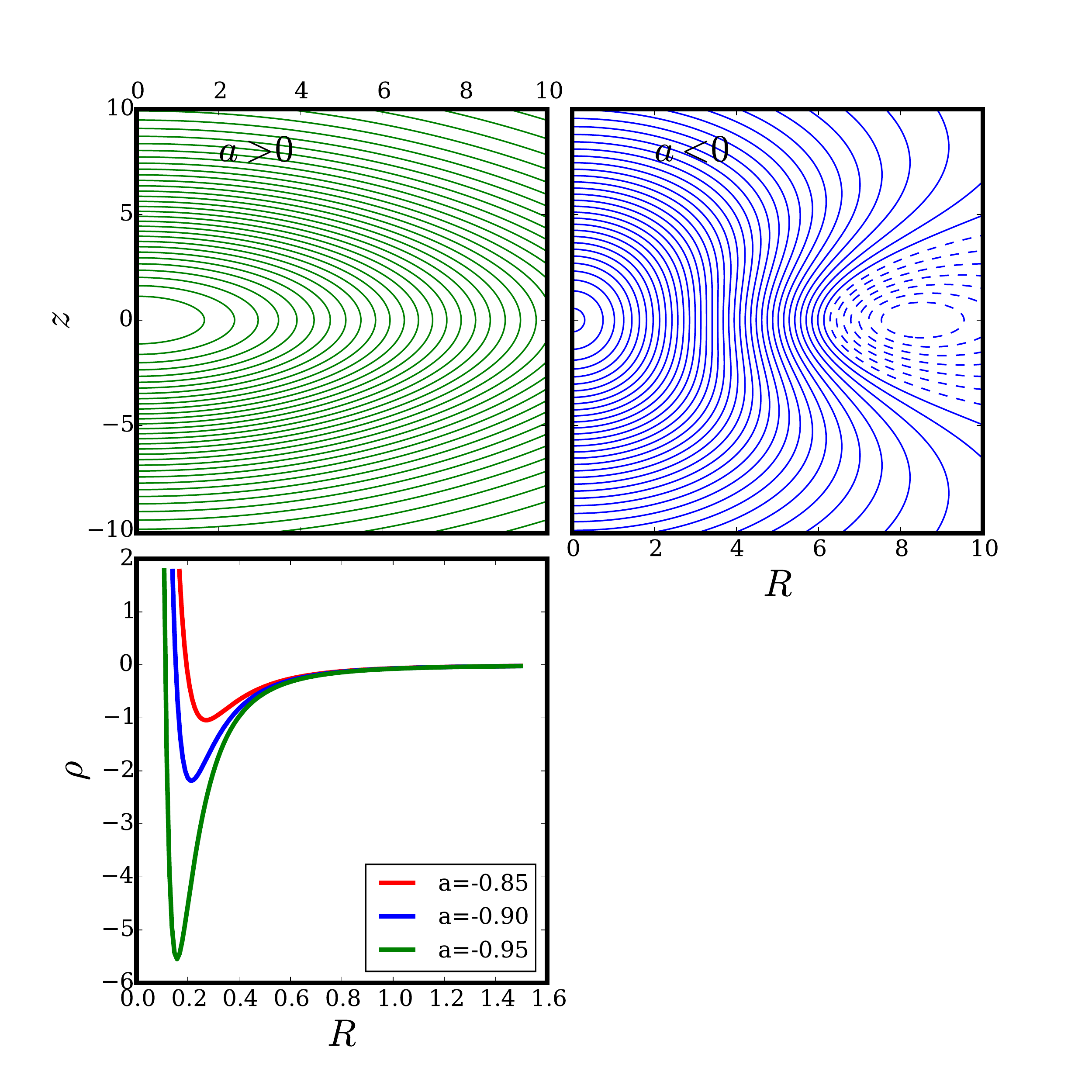}
\vspace{-10mm}
\caption{The Miyamoto Nagai `negative' (MNn) density distribution (upper panels). The left panel shows the case when all three parameters in equation \ref{DenMiya} ($M, a, b$) are positive (in this example, $M=10, a=4, b=7$). The right panel shows the case with $a<0$ ($M=10, a=-4, b=7$). In this case, there is a negative minimum in the density at $R \sim 8, z = 0$; the dashed contours represent negative density. The lower panel shows the density as a function of radius for several values of $a$ ($M=1, b=1$).}
\label{MNprofile}
\end{figure}

We make our expansion more efficient by allowing the parameter $a$ in the Miyamoto-Nagai disc (equation \ref{DenMiya}) to be negative. In this case, we should no longer think of equation \ref{DenMiya} as describing a disc, but rather a component of some more general function expansion. Note that when allowing $a < 0$, we still maintain an analytic solution to the Poisson equation.

From equation \ref{DenMiya}, we can see that there will be interesting behaviour above and below the boundary $a + b = 0$. At this point, the term in the denominator $[R^2 + (a + \sqrt{z^2+b^2})^2]^{5/2}$ produces a negative divergent density at $\rho(0,0)$. For $a + b < 0$, this divergent behaviour splits, giving two infinite minima at $R=0$, $z_c = \pm \sqrt{a^2 - b^2}$. Since such negative infinities cannot be purged through the addition of any finite number of positive MNn discs, we place the restriction $a + b > 0$ on our MNn disc models. This leads to a single negative density minimum at a point $[R_c, 0]$, where: 

\begin{equation} 
R_c = \sqrt{-\frac{(a+b)^2}{a}\left(a + 5b\right)}.
\label{eqn:rc}
\end{equation}
This is shown in Figure \ref{MNprofile}; second panel. As can be seen, the restriction $a + b > 0$ still produces a wide range of interesting behaviour that can increase the efficiency of our function expansion.

\subsection{The MNn potential-density pair}

We build our triaxial potential-density pair by summing over three sets of such MNn discs, each aligned along a symmetry plane:

\begin{equation} 
\rho(x,y,z) = \sum_{i=1}^3\sum_{j=1}^{N_i} \rho_{\rm MNn}\left(Q_i,p_i,[M_{i,j},a_{i,j},b_{i,j}]\right)
\label{eqn:MNnbasis}
\end{equation}
where $Q_1^2 = (x^2+y^2); p_1 = z$; $Q_2^2 = (y^2 +z^2); p_2 = x$; $Q_3^2 = (x^2 + z^2); p_3 = y$ and $M_{i,j},a_{i,j},b_{i,j}$ are the parameters. The total number of parameters is $3 (N_1 + N_2 + N_3)$, where $N_i$ is the number of MNn discs lying along each plane. As explained in 2.1, we apply the constraints $a_{i,j} + b_{i,j} > 0$ to ensure that there is no divergence in the density. This does not guarantee that our solution will be positive. However, since each MNn component has a single minimum that lies along $[R,0]$, we ensure global positivity by ensuring positivity along $[Q_1,0]$, $[Q_2,0]$ and $[Q_3,0]$ using Brent's method \citep{brent1973algorithms}. Note that in practice, it is very rare for the total density to go negative. For this reason, when fitting our basis to data we typically check for positivity only after the fitting is complete. This is substantially more efficient than checking `on the fly'. Our code can, however, check on the fly if necessary.

\subsection{The MCMC fit}

To test the efficiency of our new potential-density pair expansion, in this paper we fit the density profile (equation \ref{eqn:MNnbasis}) to a range of different density distributions relevant for modelling galaxies, from exponential discs to twisted triaxial ellipsoids. To perform these fits, we use the affine invariant Markov Chain Monte Carlo sampler \EMCEE\ \citep{2013PASP..125..306F}. We use 100 walkers each run for 2000 models, determining the goodness of fit from a $\chi^2$ measure between the MNn potential-density pair and the 3D target density distribution calculated on a grid. We use a grid spacing of 0.2 of the scale length, covering a range from 0.1 to 10 scale lengths. We throw out the first half of all models to ensure that our chains have `burned in'. Since we are not so interested here in the degeneracy of our model fits, we focus just on the best fitting model, as measured by the $\chi^2$ between the 3D model density distribution and the target model. With \EMCEE,\ it is straightforward to set the  constraint $a+b>0$ by setting the likelihood of the fit to $-\infty$ if the constraint is violated. Our MNn potential-density pair expansion code, including the \EMCEE\ fitting routines, is publicly available: \url{https://github.com/mdelorme/MNn}. If using the code, please cite this paper and its Github page. 

\section{Results}\label{sec:results} 

In this section, we fit our MNn expansion (equation \ref{eqn:MNnbasis}) to a range of density profiles relevant for modelling galaxies. Our aim is to test how efficient and accurate our potential-density pair is by calculating how many terms are required to give a good representation of the target density distribution. We define a `good representation' as having an error of better than 10\% over $0.1 - 10$ scale radii. This is a reasonable goal given that we are unlikely to have data of sufficient quality to measure real galactic potentials at better than this accuracy, even with Gaia quality data \citep[e.g.][]{2015arXiv150708581S}.

\subsection{Spherical target models}\label{Spherical}

We begin by considering purely spherical models. In this case, the density profile only depends on the distance to the centre of the mass distribution. We present two cases of interest. The first is the Navarro, Frenk and White (NFW) density profile:

\begin{equation}
\rho(r) = {\rho_0 \over {(r/r_s)(1+r/r_s)^2}},
\label{NFWprofile}
\end{equation}
where $r_s$ is the scale factor and $\rho_0$ is the characteristic density of the distribution. This gives a good approximation to the spherically averaged dark matter distribution in pure-dark matter structure formation simulations \citep[e.g.][]{1991ApJ...378..496D,1996ApJ...462..563N}. It is also an interesting test because the central density diverges whereas the MNn density profile has a central constant density core (see equation \ref{DenMiya} in the limit $a=0; R,z\rightarrow 0$). Thus, it should be quite challenging for our potential-density pair to reproduce equation \ref{NFWprofile}.

The second profile we consider is a `cored NFW' distribution:

\begin{equation}
\rho(r) = {\rho_0 \over {(1/r_s)\sqrt{r_c^2+r^2}(1+r^2/r_s^2)}},
\label{coredNFWprofile}
\end{equation}
where $r_c$ is the core radius. This is interesting since such dark matter density distributions appear to give a better match to real disc galaxies \citep[e.g.][]{1994ApJ...427L...1F,1994Natur.370..629M,2001ApJ...552L..23D,2011AJ....142...24O,2015AJ....149..180O}, as may be expected if dark matter is heated by bursty star formation \citep[e.g.][]{2005MNRAS.356..107R,2014Natur.506..171P,2015arXiv150804143R}.

Since the above distributions are spherically symmetric, we assume $a = 0$ and compare the MNn expansion with the target model just along the radial coordinate. In this case ($a = 0$), the MNn density profile reduces to the Plummer profile. In principle, we could add in other expansion terms like a Dehnen sphere \citep{1993MNRAS.265..250D} to make fitting cuspy profiles more efficient, but this is beyond the scope of this work. We use 50 points uniformly distributed over the range $0.1 - 10 r_s$. Using a non-uniform grid allows us to obtain a better fit at smaller radii with a correspondingly poorer fit at larger radii for a fixed number of MNn terms. This may be useful if the behaviour at large radii, where the density is very low, needs not be captured so accurately. We focus on uniform grids for the remainder of this paper. The results are shown in Figure \ref{FigSpheric}, left panels. We define the residual of the logarithm of density as follows: 

\begin{equation}
{\rm Residual} = {\rm Log}_{10}\left[\left|\frac{\rho_{\rm true}-\rho_{\rm fit}}{\rho_{\rm true}}\right|\right]
\label{residual}
\end{equation}
where $\rho_{true}$ is the value of the original density profile and $\rho_{fit}$ is the value of the MNn fitting.
As can be seen, we obtain a very good fit in both cases (at better than $\sim 10\%$ accuracy over $0.1 - 10 r_s$) with only four MNn discs. 

\subsubsection{The effect of varying the number of MNn discs}

\begin{figure}
\includegraphics[width=0.49\textwidth]{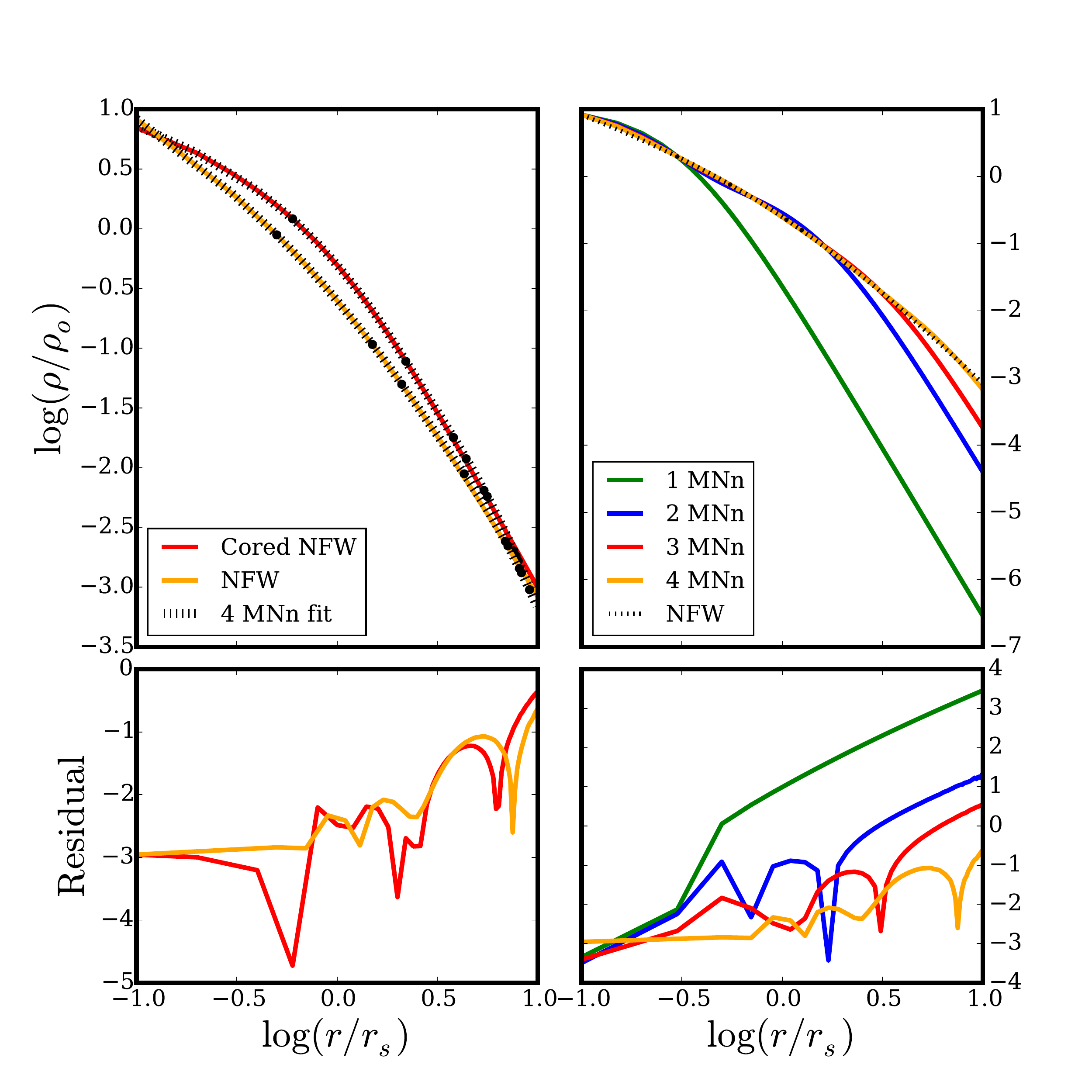}
\caption {Left panels: MNn fit (dotted lines) for two spherically symmetric density profiles, NFW (orange line) and cored NFW (red line). Right panels: fitting the NFW density profile with one, two, three and four MNn discs. The lower panels show (for this and the following figures) the residuals of the logarithm of density (see equation \ref{residual}).}
\label{FigSpheric}
\end{figure}

In Figure \ref{FigSpheric}, right panels, we compare the NFW profile with the MNn expansion, but this time varying the number of MNn discs. We perform the fitting with one, two, three and four MNn discs. As we can see clearly in the figure, the quality progressively improves as more MNn discs are added. In this paper we are primarily interested in reproducing `perfect data' and assessing whether our MNn expansion is convergent. When fitting imperfect data with our MNn expansion, we will reach a point where adding further terms to the expansion no longer improves the `Bayesian Evidence' of the fit \citep[e.g.][]{2008ConPh..49...71T}. This can be used to determine how many terms are suitable for a given data sample.

\subsection{Axisymmetric target models}\label{Exponential}

As discussed in \S\ref{sec:intro}, galactic discs are well approximated by an exponential function in $R$ and $z$:

\begin{equation}
\rho(R,z) = {\rho_0 e^{-R/R_0}e^{-z/z_0}},
\label{DobleExponential}
\end{equation}
where $R_0$ is the radial length-scale and $z_0$ is the vertical length-scale. 

\begin{figure}
\includegraphics[width=0.5\textwidth]{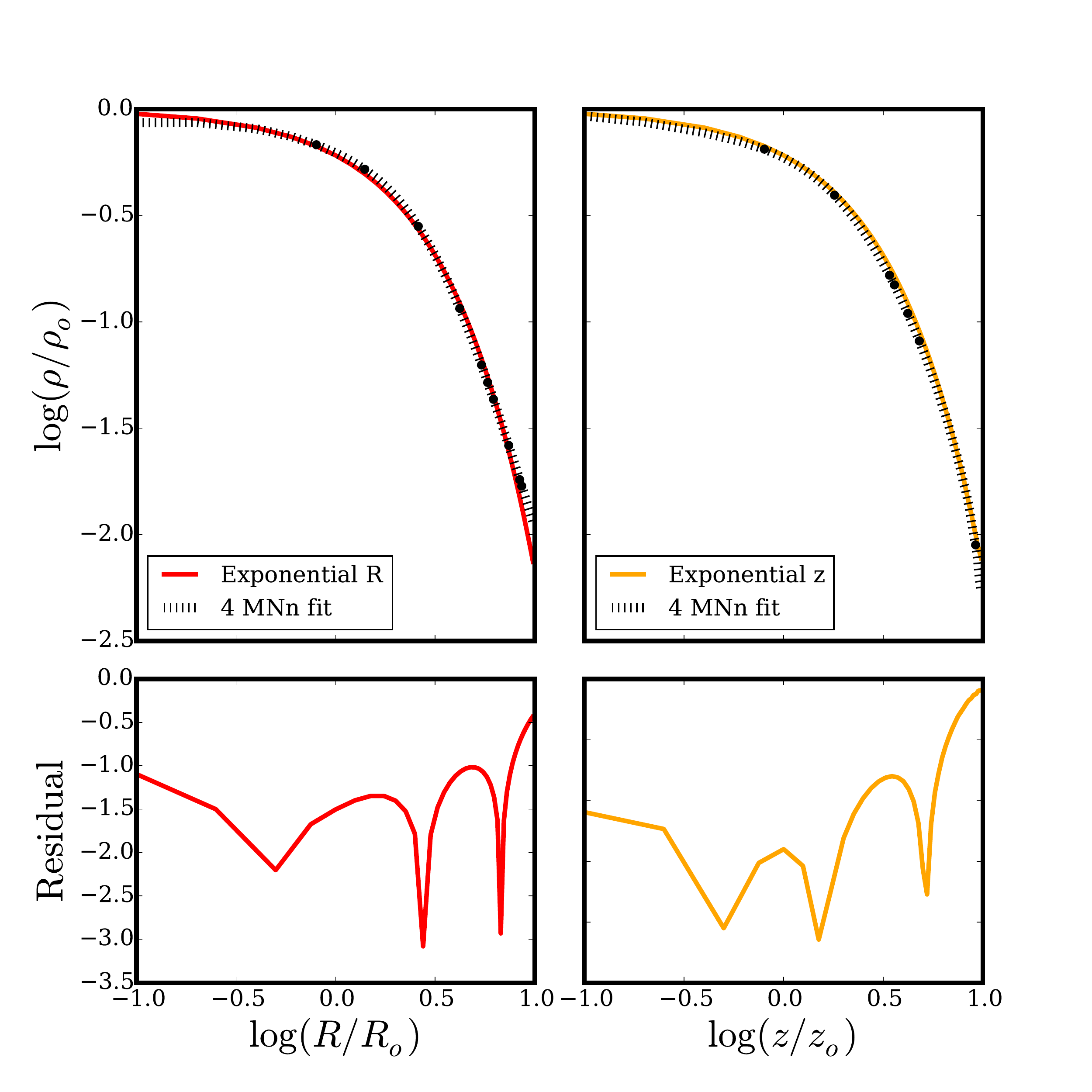}
\caption{MNn fit for the exponential disc density profile in equation \ref{DobleExponential}. The
 solid lines represent the original function and the dotted ones
 represent the MNn fit. Four MNn discs are enough to obtain a fit at better than $\sim 10\%$ accuracy over $0 < R/R_0 < 10$ and $0 < |z/z_0| < 10$.}
\label{FigExpo}
\end{figure}

We fit our MNn expansion to equation \ref{DobleExponential} by comparing the two on a 2D grid of 1600 points in $R,z$ over the range $0 < R/R_0 < 10$ and $0 < |z/z_0| < 10$. The results are shown in Figure \ref{FigExpo}. As can be seen, we obtain a fit at better than $\sim 10\%$ accuracy over $0 < R/R_0 < 10$ and $0 < |z/z_0| < 10$ with just four MNn discs. We checked the residuals over the entire grid used for the fit (here and in the following targets), and found that are comparable to those reported here along the symmetry axes. Here, we clearly benefit from the fact that the MNn disc is intrinsically axisymmetric and aligned with the target distribution.

\subsubsection{The effect of `negative $a$'}

\begin{figure}
\includegraphics[width=0.5\textwidth]{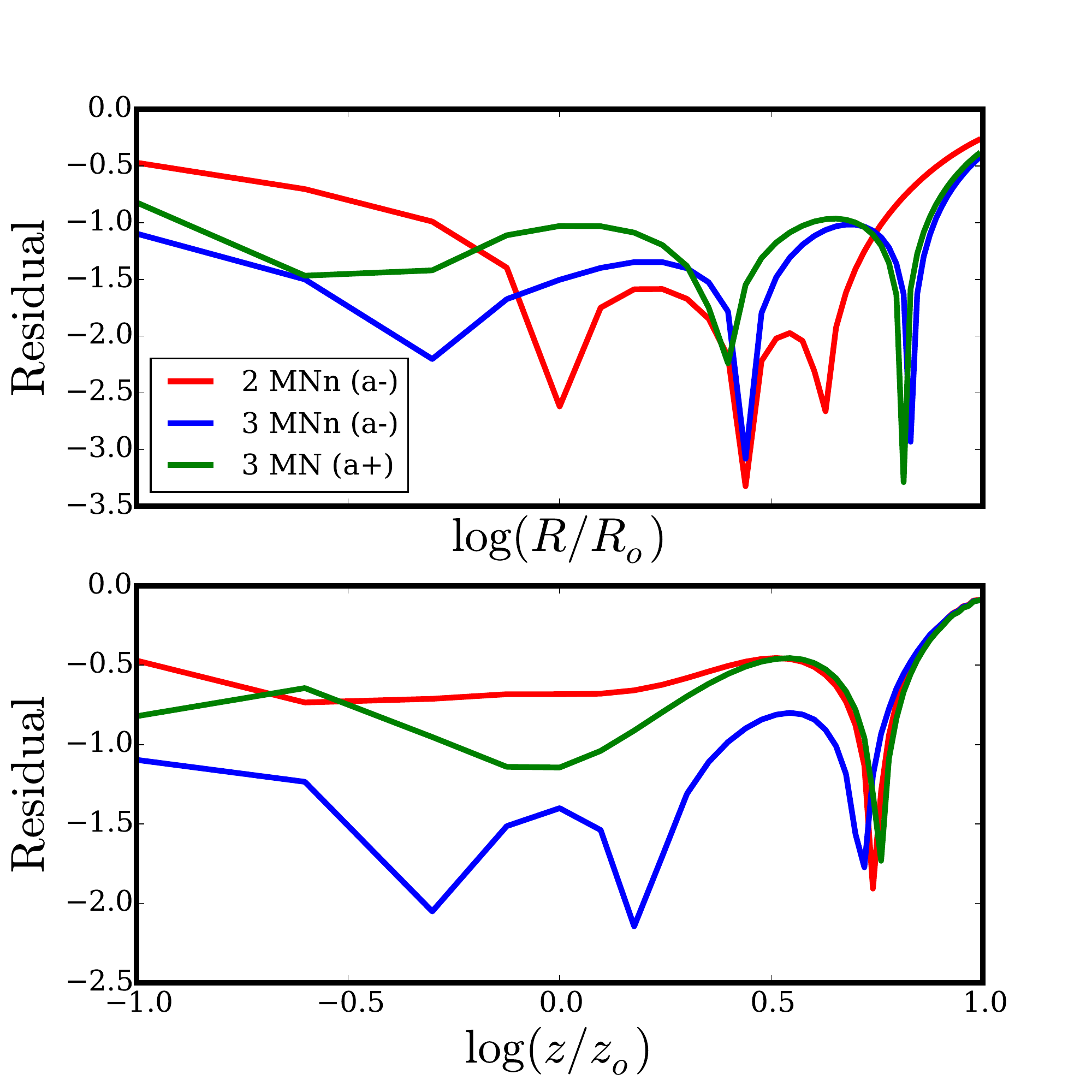}
\caption{Residuals of a double exponential density profile with the MN and MNn expansion functions. We obtain a better fit allowing $a$ to be negative with the restriction $a + b > 0$}
\label{Residual}
\end{figure}

In order to explore the effect of allowing ``negative $a$" on the fitting, we fit a double exponential density profile as follows. First, we carried out the fitting with 3 MN discs, requiring $a$ to be positive for all the three discs. Then, we performed the fitting with two and three MNn discs, that is, allowing $a$ to be negative with the restriction $a + b > 0$. Figure \ref{Residual} shows the residuals of this process. The three MNn model is the best fit out of the three, particularly with regards the vertical structure of the disc, while the two MNn and the three MN fits are about equivalent. Thus, the MNn potential-density pair gives us a substantially better fit than the MN potential-density pair for the same number of parameters.

\subsection{Triaxial target models}\label{Triaxial}

We now consider fully triaxial models for which we will need to use our full MNn expansion in equation \ref{eqn:MNnbasis}. 

\subsubsection{The triaxial NFW density profile}\label{NFWTri}

\begin{figure}
\includegraphics[width=0.5\textwidth]{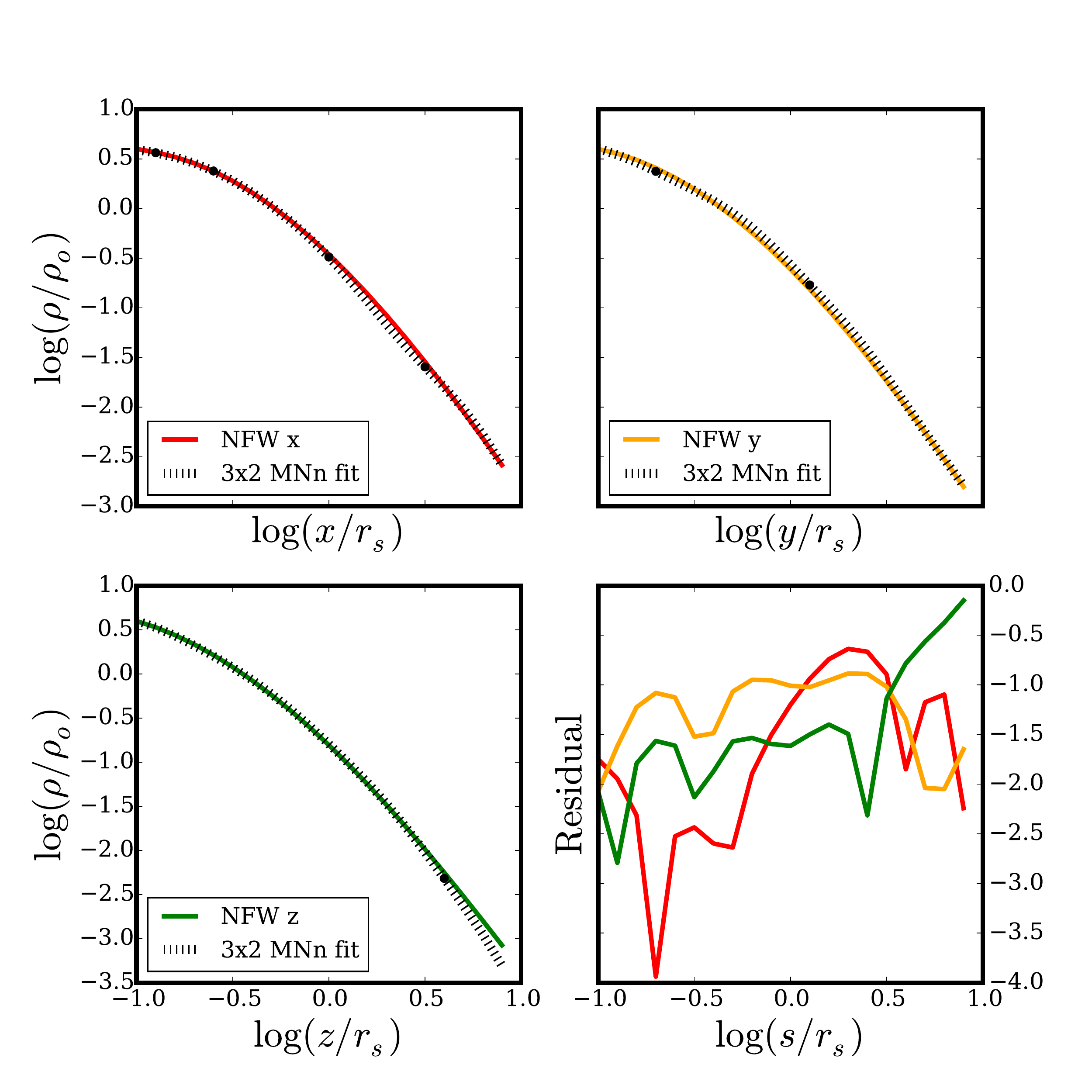}
\caption{MNn fit (dotted line) for a triaxial NFW density profile (solid lines). We require $3 \times 2$ MNn discs (two MNn discs per symmetry plane) to obtain a good fit (at better than $\sim 10$\% over the range $0.1 < s/r_s < 10$).}
\label{FigNFW}
\end{figure}

We first consider a triaxial version of the NFW density profile in equation \ref{NFWprofile}. This is interesting because dark matter halos in pure dark matter simulations are expected to be triaxial \citep[e.g.][]{2002ApJ...574..538J}, as are elliptical galaxies \citep[e.g.][]{1978ComAp...8...27B}. We consider constant ellipticity as a function of ellipsoidal radius by introducing elliptical coordinates and substituting $r$ for $s$:

\begin{equation}
s^2=\frac{x^2} {a^2}+\frac{y^2} {b^2}+\frac{z^2} {c^2},
\label{Elliptical}
\end{equation}
where $a > b > c$ are the major, intermediate and minor axis, respectively. Here, we use $a = 1.2, b = 1.0, c = 0.8$. The resulting density profile is given by:

\begin{equation}
\rho(s) = {\rho_0 \over {(s/r_s)(1+s/r_s)^2}}.
\label{TriaxialNFWprofile}
\end{equation}
We fit our full MNn expansion to equation \ref{TriaxialNFWprofile} on a 3D grid of 8000 points over the ranges $0.1 < [x,y,z]/r_s < 10$. The results are shown in Figure \ref{FigNFW}. We find that we obtain a fit at better than $\sim 10\%$ accuracy over $0.1 < s/r_s < 10$ with $3 \times 2$ MNn discs -- i.e. two MNn discs per symmetry plane.

Note that the above fit requires 18 parameters which is substantially more than the five required by equation \ref{TriaxialNFWprofile}. However, we gain two important benefits for this additional cost. Firstly, our solution has a fully analytic gravitational potential and force. Secondly, we are able with the same number of parameters to represent a much broader range of profiles than those encompassed by equation \ref{TriaxialNFWprofile}. We consider this more explicitly, next.

\subsubsection{The twisted triaxial NFW density profile}\label{Twisted}

Real dark matter halos (and elliptical galaxies) may be substantially more complex than the density profile in equation \ref{TriaxialNFWprofile}. Firstly, the triaxiality is expected to vary and `twist' with ellipsoidal radius \citep[e.g.][]{2002ApJ...574..538J}. Secondly, when considering baryon cooling and disc formation the halo is expected to become oblate and aligned with the disc -- at least out to $\sim 20$ disc scale lengths \citep[e.g.][]{1991ApJ...377..365K,1994ApJ...431..617D,2007arXiv0707.0737D}. This will lead to an inner oblate halo that becomes triaxial at large radii. There may even be evidence for such a radially varying shape from the peculiar dynamics of the Sagittarius stream stars \citep{2013ApJ...773L...4V}. To see how efficiently our expansion can represent these more complex profiles, we construct a `twisted' triaxial NFW profile as follows. We define a triaxial variable, similarly to previously:

\begin{equation}
s_T^2 = \frac{x^2} {a^2}+\frac{y^2} {b^2}+\frac{z^2} {c^2},
\label{ST}
\end{equation}
but this time we define also an axisymmetric variable:

\begin{equation}
s_A^2 = \frac{x^2} {a^2}+\frac{y^2} {b^2}+\frac{z^2} {b^2}.
\label{SA}
\end{equation}
Now, we define a variable that is axisymmetric in the centre and triaxial in the outskirts and has a smooth transition at radius $r_t$. This variable can be written as:

\begin{equation}
s' = {{r_t+s_T \over {r_t+s_A}} s_A}.
\label{SA}
\end{equation}
The twisted triaxial NFW density profile is then given by:

\begin{equation}
\rho(s') = {\rho_0 \over {(s'/r_s)(1+s'/r_s)^2}},
\label{TwistedNFWprofile}
\end{equation}
with $s'$ defined as above. Thus, when $s_T$ and $s_A$ are small compared with $r_t$, $s' \approx s_A$ and the density profile is axisymmetric. On the other hand, when $s_T$ and $s_A$ are large compared with $r_t$, $s' \approx s_T$ and the density profile is triaxial. 

\begin{figure}
\includegraphics[width=0.5\textwidth]{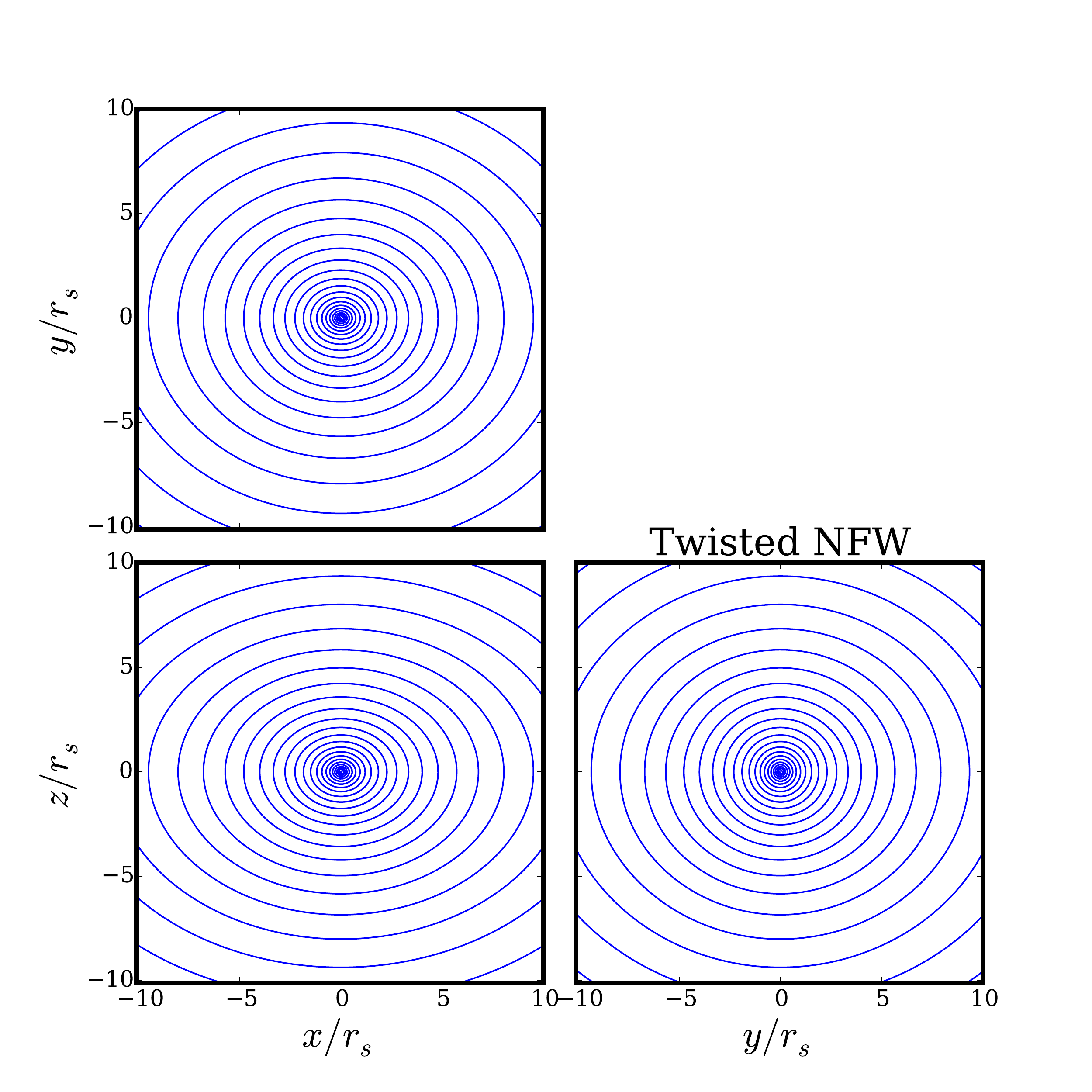}
\caption{The twisted triaxial NFW density profile. The mass distribution is axisymmetric close to the centre and triaxial in the outskirts.}
\label{TwistedProfile}
\end{figure}

\begin{figure}
\includegraphics[width=0.5\textwidth]{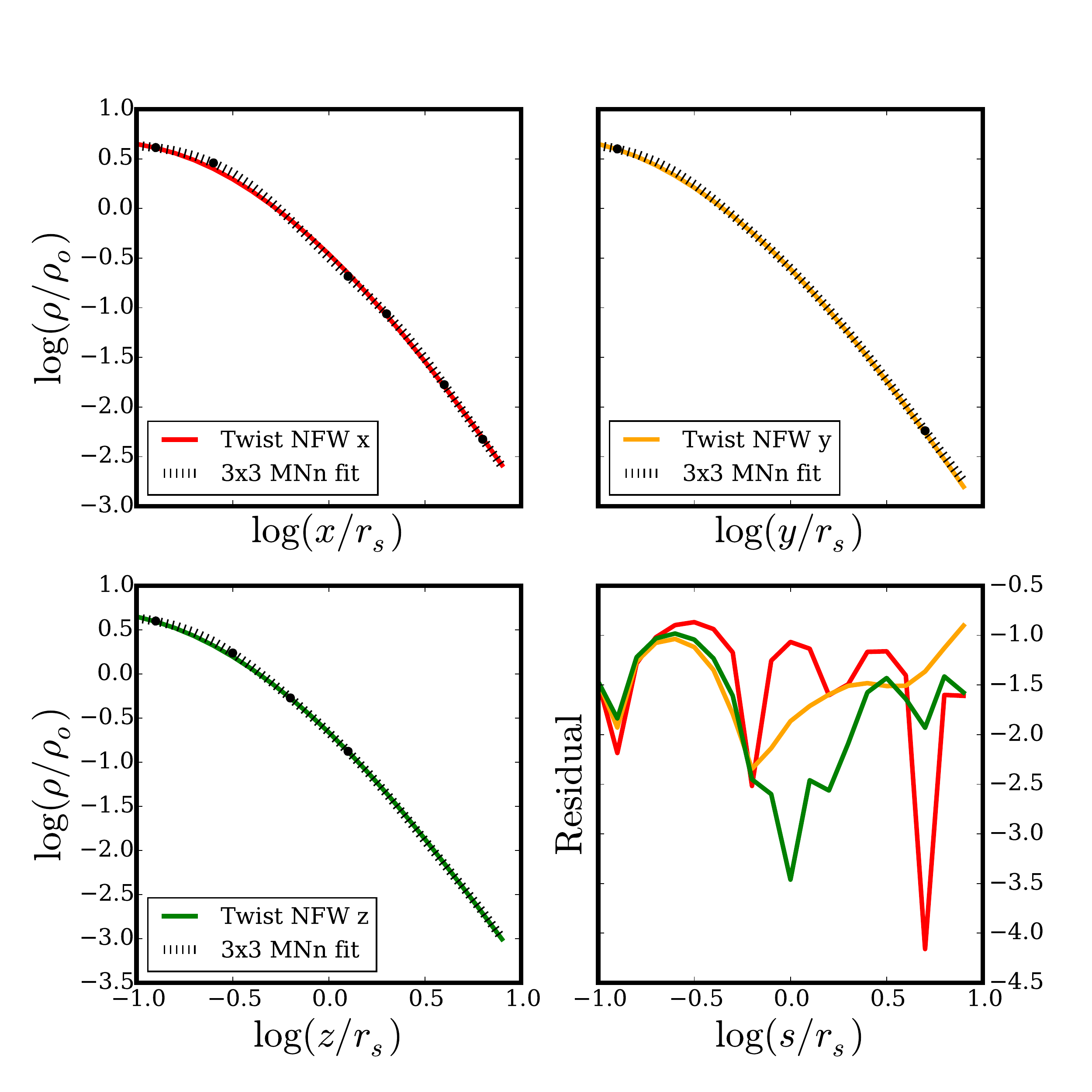}
\caption{MNn fit (dotted line) for a twisted triaxial NFW density profile (solid lines). We require $3 \times 3$ MNn discs for this more complex profile to get a good fit (better than 10\% accuracy over the range $0.1 < s'/r_s < 10$.}
\label{Twist}
\end{figure}

We fit our full MNn expansion to equation \ref{TwistedNFWprofile} on a 3D grid of 8000 points over $0.1 < [x,y,z]/r_s < 10$. We use $a = 1.2, b = 1.0, c = 0.8$, same as in the previous section. The density distribution is shown in Figure \ref{TwistedProfile} and the results of our MNn potential-density pair fit in Figure \ref{Twist}. As can be seen, we now require $3 \times 3$ sets of MNn discs (i.e. 27 parameters) to obtain a fit at better than 10\% accuracy over $0.1 < s'/r_s < 10$.

\subsection{The potential and the force field}\label{Potential}

So far, we have compared the original density profile with the one we obtain with our potential-density pair. It is interesting also to compare the potential generated by these density profiles and the force field as well. An important advantage of our potential-density pair is that the potential and the force field can be calculated analytically which is not the case, for instance, for the triaxial and twisted NFW profiles which have to be calculated numerically.
 
\begin{figure}
\includegraphics[width=0.5\textwidth]{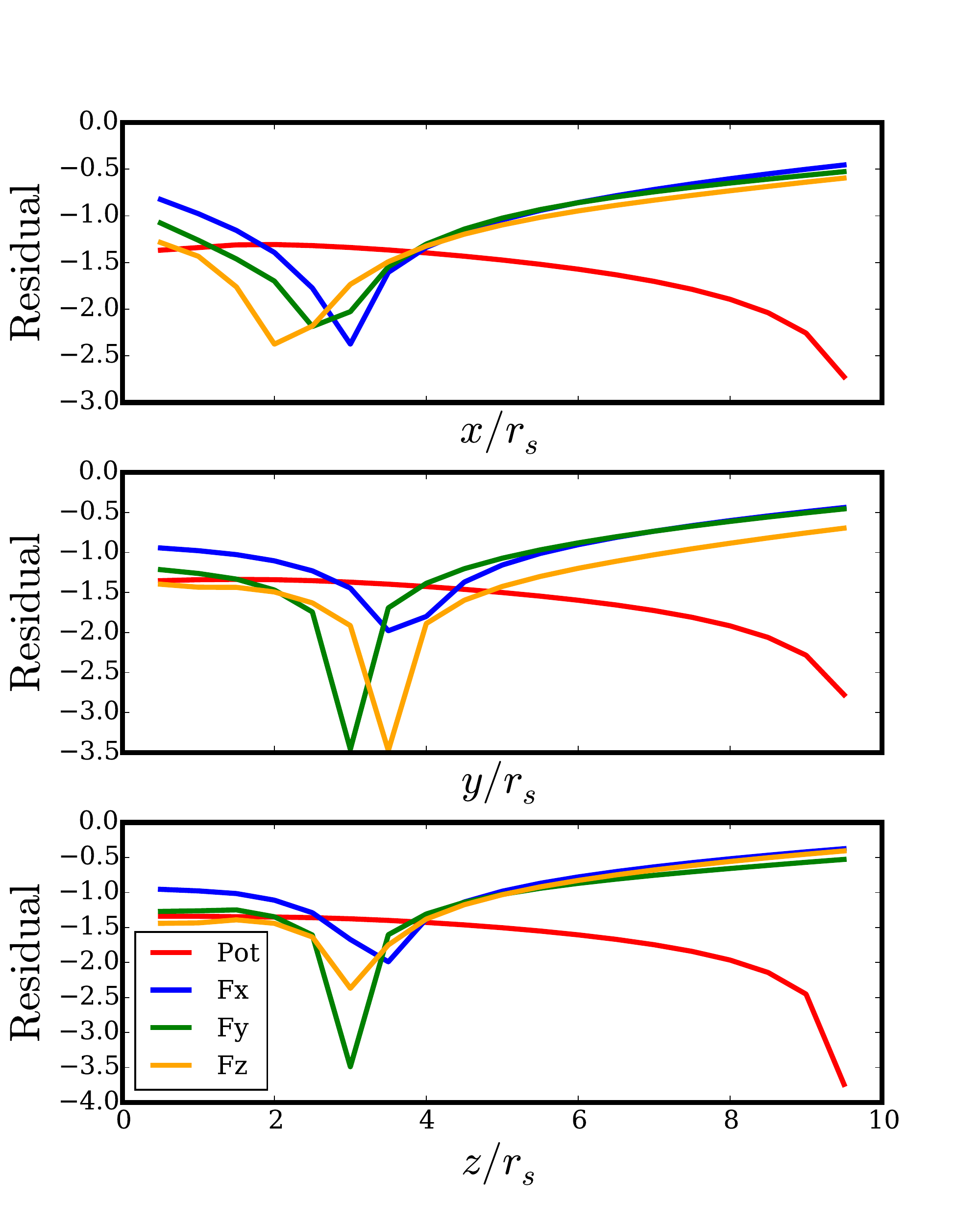}
\caption{Residuals of the potential and force field for a $3 \times 2$ MNn fit to the triaxial NFW density profile.}
\label{Force}
\end{figure}

We now compare the potential and the force field generated by a triaxial NFW density profile with those generated by our potential-density pair. We first calculate the potential generated by the original function, using the Chandrasekhar formula \citep{chandrasekhar1969ellipsoidal}:

\begin{eqnarray}
  \Phi(x,y,z)=2\pi G abc\rho_0 r_s^2 \hspace{3.5cm} \nonumber \\ 
\hspace{0.3cm} \times \int_0^\infty \frac
       {s(\tau)}{r_s+s(\tau)} \frac{d\tau } {\sqrt
         {(a^2+\tau)(b^2+\tau)(c^2+\tau)}}.
\end{eqnarray}
This is an improper integral over the variable $\tau$, where

\begin{equation}
 s(\tau)^2=\frac{x^2} {a^2+\tau}+\frac{y^2} {b^2+\tau}+\frac{z^2} {c^2+\tau}.
\label{stau}
\end{equation}
Once we have calculated the potential generated by the original density profile, now we calculate the potential generated by our superposition of MNn discs. This is just the sum of the potential generated by each single MNn disc that follows analytically from equation \ref{PotMiya}. We compare our MNn potential and force field with the true solution in Figure \ref{Force} for the case of a triaxial NFW profile.

\section{Conclusions}\label{conclusions}

We have presented a novel positive potential-density pair expansion for modelling galaxies, based on the Miyamoto-Nagai (MN) disc. The key novelties of our method are: (i) we use three sets of such discs aligned along each symmetry axis, allowing us to model triaxial and twisted triaxial systems; (ii) we increase the efficiency of our expansion by allowing the scale length parameter of each disc $a$ to be negative; (iii) we introduce a constraint equation that ensures that the total density is everywhere positive; and (iv) we allow arbitrarily many discs in our expansion, allowing us to accurately model a wide range of density distributions relevant for modelling galaxies. We find that at better than 10\% accuracy in our density reconstruction, we can represent a radial and vertical exponential disc over $0.1-10$ scale lengths/heights with 4 MNn discs; an NFW profile over $0.1-10$ scale lengths with 4 MNn discs; and a twisted triaxial NFW profile with 3 MNn discs per symmetry axis. Our potential-density pair is efficient, fully analytic, and well-suited to reproducing the density distribution and gravitational potential of galaxies from discs to ellipsoids.

\section*{Acknowledgments}

ARN would like to acknowledge support from CONACyT complementary postdoctoral fellowship, proposal 238583. JIR would like to acknowledge support from STFC consolidated grant ST/M000990/1 and the MERAC foundation. LA acknowledges support from UNAM/DGAPA PAPIIT grant IG100115. We would like to thank Walter Dehnen and the anonymous referee for useful feedback that improved the paper.

\bibliographystyle{mnras}
\bibliography{refs}

\end{document}